\documentclass[12pt,a4wide,epsf]{article}
\usepackage{epsf}
\usepackage{graphicx}
 \newcommand{\insertplot}[5]{\begin{figure}
 \hfill\hbox to 0.05in{\vbox to #5in{\vfill
 \inputplot{#1}{#4}{#5}}\hfill}
 \hfill\vspace{-.1in}
 \caption{#2}\label{#3}
 \end{figure}}
 \newcommand{\inputplot}[3]{
 \special{ps: plotfile #1}
}
 \newcommand{\vphi}{\varphi}
\newcounter{fig}   \newcommand{\lbfig}[1]{\refstepcounter{fig}
\label{#1} }

\begin{document}
\title{Rotating Black Holes with Monopole Hair}
\author{
{\bf Burkhard Kleihaus and Jutta Kunz}\\
Institut f\"ur Physik, Universit\"at Oldenburg, Postfach 2503\\
D-26111 Oldenburg, Germany\\
\\
{\bf Francisco Navarro-L\'erida}\\
Dept.~de F\'{\i}sica At\'omica, Molecular y Nuclear, Ciencias F\'{\i}sicas\\
Universidad Complutense de Madrid, E-28040 Madrid, Spain}

\date{\today}

\maketitle

\begin{abstract}
We study rotating black holes in Einstein-Yang-Mills-Higgs theory.
These black holes emerge from static black holes with monopole hair
when a finite horizon angular velocity is imposed.
At critical values of the horizon angular velocity and the horizon radius,
they bifurcate with embedded Kerr-Newman black holes. 
The non-Abelian black holes possess an electric dipole moment,
but no electric charge is induced by the rotation.
We deduce that gravitating regular monopoles possess
a gyroelectric ratio $g_{\rm el}=2$.

\end{abstract}

\vfill\eject

\section{Introduction}

SU(2) Einstein-Yang-Mills-Higgs (EYMH) theory, with the Higgs field
in the adjoint representation,
possesses globally regular gravitating magnetic monopole solutions
\cite{gmono1,gmono2}.
For small gravitational constant,
a branch of gravitating monopole solutions emerges smoothly
from the corresponding flat space solution,
the 't Hooft-Polyakov monopole \cite{mono}.
With increasing gravitational constant,
the mass of the gravitating monopole solutions decreases.
The branch of gravitating monopole solutions
extends up to a maximal value of the gravitational constant.
For vanishing Higgs self-coupling constant, the monopole branch then
bends backwards and merges with the branch of 
extremal Reissner-Nordstr\"om (RN) solutions 
at a critical value of the coupling constant.

Magnetically charged EYMH black hole solutions
emerge from the globally regular magnetic monopole solutions
when a finite regular event horizon is imposed \cite{gmono1,gmono2}. 
Consequently,
they have been characterized as ``black holes within magnetic monopoles''
\cite{gmono1}.
Distinct from embedded RN black holes with unit magnetic charge,
they possess non-Abelian monopole hair.
For small values of the gravitational constant,
the black hole solutions merge into non-extremal RN solutions
at a critical value of the horizon radius.
For larger values of the gravitational constant,
the black hole solutions show a critical behaviour
analogous to the globally regular solutions,
and bifurcate with extremal RN solutions.

In this letter we demonstrate, 
that black hole solutions with non-Abelian monopole hair
possess rotating generalizations.
Previously rotating non-Abelian black hole solutions were
known only in Einstein-Yang-Mills(-dilaton) theory \cite{vs,kkrot,kkrotd}.

We investigate the domain of existence of these
rotating non-Abelian black hole solutions.
We observe, that at critical values of the horizon radius 
and the horizon angular velocity,
the rotating non-Abelian black hole solutions
bifurcate with rotating Abelian black hole solutions,
the Kerr-Newman (KN) black holes.

In contrast to non-Abelian black holes,
regular monopole solutions cannot rotate \cite{heusler,radu}.
Still, monopole solutions in flat space are known to possess a
gyroelectric ratio of two \cite{gel}.
We here deduce the gyroelectric ratio for gravitating
monopole solutions.

We review the EYMH action in section 2. We present
the ans\"atze for the metric and the fields in section 3, 
and the boundary conditions in section 4.
We address the properties of the solutions in section 5
and present our numerical results in section 6.

\section{Action}

We consider the SU(2) EYMH action
in the limit of vanishing Higgs potential,
\begin{eqnarray}
S &=& \int \left [ \frac{R}{16\pi G}
  - \frac{1}{2} {\rm Tr} \left(F_{\mu\nu} F^{\mu\nu}\right)
 -\frac{1}{4} {\rm Tr} \left( D_\mu \Phi D^\mu \Phi \right)
 \right ] \sqrt{-g}\, d^4x
\ \label{action} \end{eqnarray}
with curvature scalar $R$,
SU(2) field strength tensor
\begin{equation}
F_{\mu \nu} =
\partial_\mu A_\nu -\partial_\nu A_\mu + i e \left[A_\mu , A_\nu \right]
\ , \label{fmn} \end{equation}
gauge potential $A_\mu = 1/2 \tau^a A_\mu^a$,
gauge covariant derivative
\begin{equation}
D_\mu = \nabla_\mu +ie [ A_\mu, \cdot \ ]
\ , \label{Dmu} \end{equation}
and Higgs field $\Phi = \tau^a \Phi^a$;
$G$ is Newton's constant, and $e$ is the gauge coupling constant.
We impose a Higgs field vacuum expectation value $v$.

Variation of the action Eq.~(\ref{action}) with respect to the metric
$g_{\mu\nu}$, the gauge potential $A_\mu^a$, and the Higgs field $\Phi^a$
leads to the Einstein equations and the matter field equations.

\section{Ans\"atze}

We consider stationary, axially symmetric black hole space times
with Killing vector fields $\xi=\partial_t$ and $\eta=\partial_{\varphi}$.
The Lewis-Papapetrou form of the metric reads in isotropic coordinates
\cite{kkrot}
\begin{equation}
ds^2 = -fdt^2+\frac{m}{f}\left[dr^2+r^2 d\theta^2\right] 
       + \frac{l r^2 \sin^2\theta}{f}
          \left[d\varphi-\frac{\omega}{r}dt\right]^2 \  
\ .  \label{am} \end{equation}
The gauge potential is parametrized by \cite{kkrot}
\begin{equation}
A_\mu dx^\mu
  =   \left( B_1 \frac{\tau_r}{2e} + B_2 \frac{\tau_\theta}{2e} \right) dt
+ A_\vphi (d\vphi-\frac{\omega}{r} dt)
+\left(\frac{H_1}{r}dr +(1-H_2)\, d\theta \right)\frac{\tau_\vphi}{2e}
\ , \label{a1} \end{equation}
with
\begin{equation}
A_\vphi=   -\sin\theta\left[H_3 \frac{\tau_r}{2e}
            +(1-H_4) \frac{\tau_\theta}{2e}\right]
\ , \label{a3} \end{equation}
and the Higgs field by
\begin{equation}
\Phi =v \left( \Phi_1 \tau_r + \Phi_2 \tau_\theta \right)
\ , \label{a2} \end{equation}
where the symbols $\tau_r$, $\tau_\theta$, and $\tau_\vphi$
denote the dot products of the Cartesian vector of Pauli matrices,
$\vec \tau = ( \tau_x, \tau_y, \tau_z) $,
with the spherical spatial unit vectors.
All functions depend on $r$ and $\theta$.

The ansatz is form-invariant under Abelian gauge transformations $U$
\cite{kkrot}
\begin{equation}
 U= \exp \left({\frac{i}{2} \tau_\varphi \Gamma(r,\theta)} \right)
\ .\label{gauge} \end{equation}
With respect to this residual gauge degree of freedom
we choose the gauge fixing condition
$r\partial_r H_1-\partial_\theta H_2 =0$
\cite{kkrot}.

\section{Boundary Conditions}

The event horizon 
resides at a surface of constant radial coordinate, $r=r_{\rm H}$,
and is characterized by the condition $f(r_{\rm H})=0$ \cite{kkrot}.
The Killing vector field $\chi = \xi + {\Omega} \eta$
is orthogonal to and null on the horizon,
where $\Omega$ denotes the horizon angular velocity \cite{wald}. 

At the horizon, we impose the boundary conditions \cite{kkrot}
$f=m=l=0$, 
$\omega=\omega_{\rm H}$,
$H_1=0$,
$\partial_r H_2= \partial_r H_3= \partial_r H_4= 0$,
$B_1=\Omega \cos \theta$, $B_2 = - \Omega \sin \theta$,
($\Omega = {\omega_{\rm H}}/{r_{\rm H}}$,) 
$\partial_r \Phi_1= \partial_r \Phi_2=0$.

At infinity we impose the boundary conditions \cite{kkrot}
$f=m=l=1$, $\omega=0$,
$H_1=H_2=H_3=H_4=0$, $B_1=B_2=0$, 
$\Phi_1=1$, $\Phi_2=0$.

On the symmetry axis, we impose \cite{kkrot}
$\partial_\theta f = \partial_\theta m = \partial_\theta l =
\partial_\theta \omega = 0$,
$H_1=H_3=0$,
$\partial_\theta H_2 =\partial_\theta H_4=0$,
$\partial_\theta B_1 =0$, $B_2=0$,
$\partial_\theta \Phi_1 =0$, $\Phi_2=0$.
Regularity further requires $m=l$ and $H_2=H_4$ on the symmetry axis.

\section{Black Hole Properties}

Let us introduce the dimensionless coordinate $x$ 
and the dimensionless coupling constant $\alpha$ \cite{gmono2},
\begin{equation}
x = \frac{e\alpha}{\sqrt{4 \pi G}} \, r \ , \ \ \
\alpha = \sqrt{4 \pi G} v \   \ \ \ 
\ . \end{equation}
The dimensionless mass $M$ and angular momentum $J$ of the black holes
are obtained from the asymptotic expansion of the metric functions
\begin{equation}
 f \rightarrow 1-\frac{2M}{x} \ , \ \ \ \omega \rightarrow \frac{2 J}{x^2}
\ . \end{equation}
The asymptotic expansion of the gauge fields shows, that
the black holes carry unit magnetic charge, $P=1$,
and no electric charge, $Q=0$, 
like the static (non-dyonic) EYMH black holes \cite{gmono1,gmono2}.
In contrast, rotating black hole solutions 
of Einstein-Yang-Mills theory 
have an electric charge induced by the rotation \cite{vs,kkrot}.
The rotating black holes do possess an electric dipole moment $\mu_{\rm el}$,
as seen in the asymptotic expansion of
the dimensionless radial component of the electric field
\begin{equation}
F_{tr} = 2 \mu_{\rm el} \frac{\cos\theta}{x^3}  \frac{\tau_r}{2} + ...
\ . \end{equation}
The dipole moment $\mu_{\rm el}$ defines the
gyroelectric ratio $g_{\rm el}$
\begin{equation}
\mu_{\rm el} = g_{\rm el} \frac{P }{2 M} J 
\ , \end{equation}
i.e., $g_{\rm el}=2 \mu_{\rm el}/a$, since $P=1$, and $a=J/M$ is
the specific angular momentum.
The full asymptotic expansion will be given elsewhere \cite{long}.

The area parameter $x_\Delta$ of the black holes
is defined via the black hole horizon area $A=4 \pi x_\Delta^2$.
Their surface gravity $\kappa$ is obtained from 
the Killing vector $\chi$ \cite{wald}
\begin{equation}
\kappa^2 = - \frac{1}{4} (\nabla_\mu \chi_\nu)(\nabla^\mu \chi^\nu)
\ . \label{sgwald} \end{equation}
As required by the zeroth law of black hole mechanics,
$\kappa$ is constant at the horizon.

\section{Results}

The numerical calculations are performed in terms of
compactified coordinates $\bar x = 1-(x_{\rm H}/x)$,
and employ a Newton-Raphson scheme 
\cite{kkrot,fidisol}.

Let us briefly recall the domain of existence and
the critical behaviour of the static monopole and black hole solutions.
For vanishing Higgs potential, the branch of monopole solutions 
extends from $\alpha=0$, the flat space limit,
to the maximal value $\alpha_{\rm max}\approx 1.4$,
and then bends backwards and merges with 
the branch of extremal RN solutions at $\alpha_{\rm cr}$
\cite{gmono2}.

For a given value of $\alpha$, 
a branch of non-Abelian black holes 
emerges from the corresponding regular solution.
This branch extends from zero to a maximal value of the isotropic
horizon radius $x_{\rm H}^{\rm max}(\alpha)$, 
and then bends backwards and merges with the branch of RN solutions
at $x_{\rm H, cr}(\alpha)$ \cite{gmono2}.
When $\alpha < \sqrt{3}/2$,
the bifurcation takes place at a non-extremal RN black hole, 
i.e., $x_{\rm H, cr}(\alpha)>0$. 
In contrast, when $\alpha \ge \sqrt{3}/2 $,
the bifurcation takes place at an extremal RN black hole,
i.e., $x_{\rm H, cr}(\alpha)=0$.

This is illustrated in Fig.~\ref{f-1}, where we show
the dependence of the mass on the isotropic horizon radius $x_{\rm H}$
for static non-Abelian black hole solutions
for $\alpha=0.3$, $\sqrt{3}/2$ and $1$, 
and also for the corresponding RN solutions.
(Note, that the analyses of Refs.~\cite{gmono1,gmono2}
are in terms of the area parameter $x_{\Delta}$.)

\begin{figure}[h!]
\lbfig{f-1}
\parbox{\textwidth}
{\centerline{
\mbox{
\epsfysize=8.0cm
\includegraphics[width=90mm,angle=0,keepaspectratio]{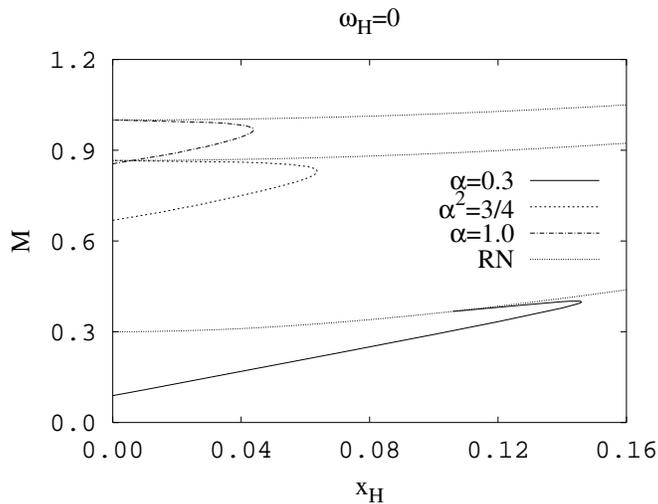}
} } }
\caption{
{\small
The dimensionless mass $M$ versus the
isotropic horizon radius $x_{\rm H}$
of static non-Abelian black holes
for $\alpha=0.3$, $\sqrt{3}/2$, and $1$. 
The mass of the corresponding RN black holes is also shown.
}}
\end{figure}

To address the domain of existence and 
the critical behaviour of rotating non-Abelian black hole solutions,
we now consider the dependence of the black hole solutions
on the horizon angular velocity parameter $\omega_{\rm H}$ 
for fixed isotropic horizon radius $x_{\rm H}$ and 
fixed coupling constant $\alpha$.

When $\omega_{\rm H}$ is increased from zero,
a branch of rotating non-Abelian black hole solutions
with horizon radius $x_{\rm H}$ 
emerges from the corresponding (lower mass) static solution.
This branch extends up to a maximal value 
$\omega_{\rm H}^{\rm max}(x_{\rm H},\alpha)$,
and then bends backwards. 

For $\alpha < \sqrt{3}/2$, and 
$x_{\rm H} < x_{\rm H, cr}(\alpha)$,
this branch merges with the branch of KN black holes
at a critical value $\omega_{\rm H, cr}(x_{\rm H},\alpha)$.
This is illustrated in Fig.~\ref{f-2},
where we show the mass of the non-Abelian black holes
as a function of $\omega_{\rm H}$ for $\alpha=0.3$ and
$x_{\rm H}=0.07$ and $x_{\rm H}=0.1$
($x_{\rm H, cr}(0.3) \approx 0.106$).
The critical value $\omega_{\rm H, cr}(x_{\rm H},\alpha)$ 
moves from the upper Abelian branch, 
where it is located for $x_{\rm H}=0.07$,
to the lower Abelian branch, when $x_{\rm H}=0.1$.

As $x_{\rm H}$ is increased further,
the critical value $x_{\rm H,cr}(\alpha)$ is passed, beyond which
two static non-Abelian solutions exist.
For the rotating black hole solutions this is reflected in the fact that,
for $x_{\rm H,cr}(\alpha) < x_{\rm H} < x_{\rm H}^{\rm max}(\alpha)$,
the branch of non-Abelian solutions does not bifurcate
with the branch of KN black hole solutions anymore,
but instead leads back to the upper static non-Abelian solution.
This is illustrated in Fig.~\ref{f-2} for $x_{\rm H}=0.14$.
As $x_{\rm H}$ approaches $x_{\rm H}^{\rm max}(\alpha)$ finally,
the branch of rotating non-Abelian black hole solutions shrinks
to zero size. 

\begin{figure}[h!]
\lbfig{f-2}
\parbox{\textwidth}
{\centerline{
\mbox{
\epsfysize=8.0cm
\includegraphics[width=90mm,angle=0,keepaspectratio]{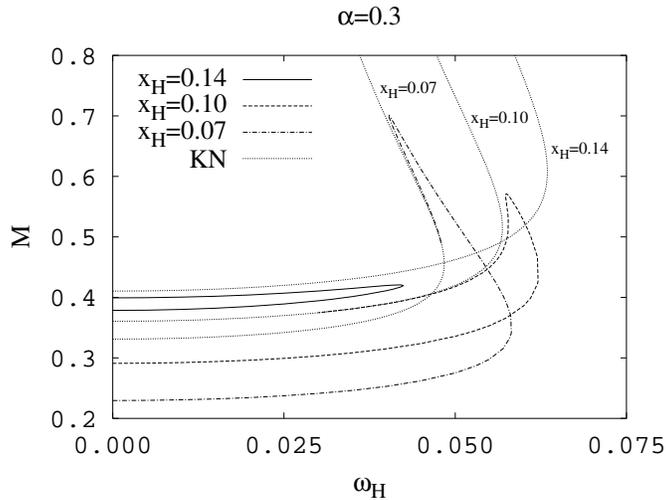}
 } } }
\caption{
{\small
The dimensionless mass $M$ versus the 
horizon angular velocity parameter $\omega_{\rm H}$ 
of non-Abelian black holes
with horizon radius $x_{\rm H}=0.07$, $0.1$, $0.14$,
and coupling constant $\alpha=0.3$.
The mass of the corresponding KN black holes is also shown.
}}
\end{figure}

As in the latter case, 
there are two static non-Abelian black hole solutions
for $\alpha > \sqrt{3}/2$,
and $x_{\rm H} < x_{\rm H}^{\rm max}(\alpha)$.
Consequently, no bifurcations with KN black holes occur, 
but the branch of rotating non-Abelian black hole solutions 
leads from the lower static non-Abelian solution
back to the upper static non-Abelian solution.
This is seen in Fig.~\ref{f-3} for $\alpha=1$
and $x_{\rm H}=0.02$ and $x_{\rm H}=0.04$.

\begin{figure}[h!]
\lbfig{f-3}
\parbox{\textwidth}
{\centerline{
\mbox{
\epsfysize=8.0cm
\includegraphics[width=90mm,angle=0,keepaspectratio]{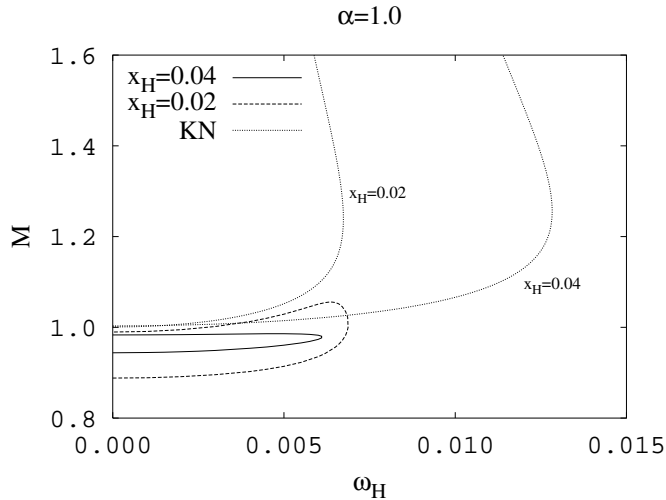}
} } }
\caption{
{\small
Same as Fig.~\ref{f-2} for $x_{\rm H}=0.02$ and $0.04$ 
and $\alpha=1$.
}}
\end{figure}

We next consider
the dependence of the rotating non-Abelian black hole solutions
on the horizon radius $x_{\rm H}$ 
for fixed horizon angular velocity parameter $\omega_{\rm H}$ and
fixed coupling constant $\alpha$.
For finite values of $\omega_{\rm H}$, 
rotating non-Abelian black hole solutions
exist only for $x_{\rm H} > x_{\rm H}^{\rm min}(\omega_{\rm H},\alpha)$.
Regular solutions are reached only in the static limit.
The ansatz Eqs.~(\ref{am})-(\ref{a2}) does not allow for
regular rotating solutions \cite{radu}.

For $\alpha < \sqrt{3}/2$,
and small $\omega_{\rm H}$,
two branches of rotating non-Abelian black hole solutions
bifurcate at $x_{\rm H}^{\rm min}(\omega_{\rm H},\alpha)$.
The lower branch follows closely the static branch, i.e., it reaches a 
maximal value $x_{\rm H}^{\rm max}(\omega_{\rm H},\alpha)$, bend backwards
and merges with the lower branch of KN black holes.
Likewise, the upper branch bifurcates with the upper
branch of KN black holes.

As $\omega_{\rm H}$ is increased, the two bifurcation values
move closer together, until they meet at a critical
value of $\omega_{\rm H}$.
When $\omega_{\rm H}$ is increased further,
the rotating non-Abelian black hole solutions no longer
bifurcate with Abelian black holes. Subsequently, the region of
existence of rotating non-Abelian black hole solutions
shrinks to zero size, as $\omega_{\rm H}$ approaches its
maximal value.
This is illustrated in Fig.~\ref{f-4}
where we show the mass of the rotating non-Abelian black holes
as a function of $x_{\rm H}$ for $\alpha=0.3$ and
$\omega_{\rm H}=0.03$, $0.05$, $0.055$, and $0.0625$.

\begin{figure}[h!]
\lbfig{f-4}
\parbox{\textwidth}
{\centerline{
\mbox{
\epsfysize=8.0cm
\includegraphics[width=90mm,angle=0,keepaspectratio]{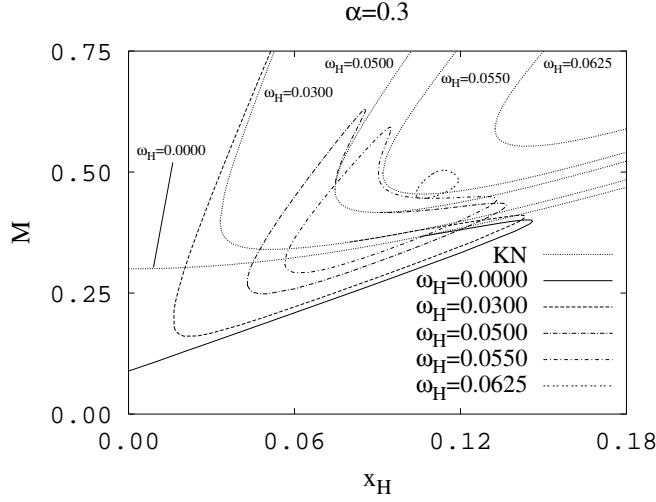}
} } }
\caption{
{\small
The dimensionless mass $M$ versus the
horizon radius $x_{\rm H}$
of non-Abelian black holes
with horizon angular velocity parameter $\omega_{\rm H}=0.03$,
$0.05$, $0.055$, $0.0625$,
and coupling constant $\alpha=0.3$.
The mass of the corresponding KN black holes is also shown,
together with mass of the static non-Abelian and RN black holes.
}}
\end{figure}

For $\alpha > \sqrt{3}/2$,
again we do not observe bifurcations with KN black holes.
As $\omega_{\rm H}$ is increased towards its maximal value
the region of existence of rotating non-Abelian black hole solutions
shrinks to zero size.
This is illustrated in Fig.~\ref{f-5}
for $\omega_{\rm H}=0.005$ and $0.0065$ and $\alpha=1$.

\begin{figure}[h!]
\lbfig{f-5}
\parbox{\textwidth}
{\centerline{
\mbox{
\epsfysize=8.0cm
\includegraphics[width=90mm,angle=0,keepaspectratio]{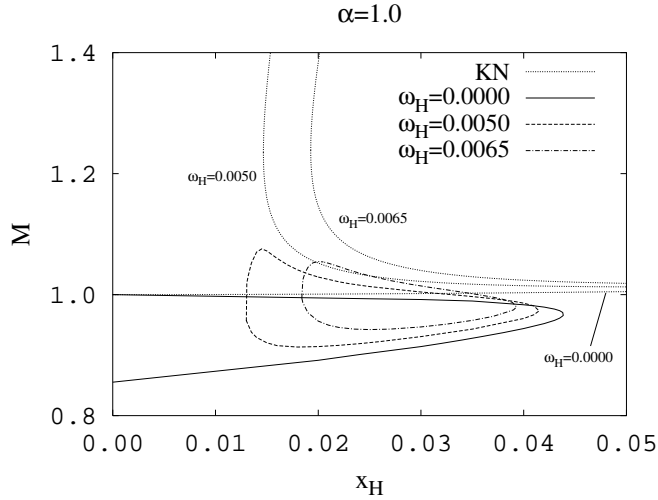}
} } }
\caption{
{\small
Same as Fig.~\ref{f-4} for $\omega_{\rm H}=0.005$ and $0.0065$
and $\alpha=1$.
}}
\end{figure}

Turning to further properties of these
rotating non-Abelian black holes,
we exhibit in Fig.~\ref{f-6} the specific angular momentum $a$
as a function of $\omega_{\rm H}$ 
for the set of solutions whose mass is shown in Fig.~\ref{f-2}.
Starting at zero from the static solution,
the angular momentum then shows a behaviour analogous to the mass.

\begin{figure}[h!]
\lbfig{f-6}
\parbox{\textwidth}
{\centerline{
\mbox{
\epsfysize=8.0cm
\includegraphics[width=90mm,angle=0,keepaspectratio]{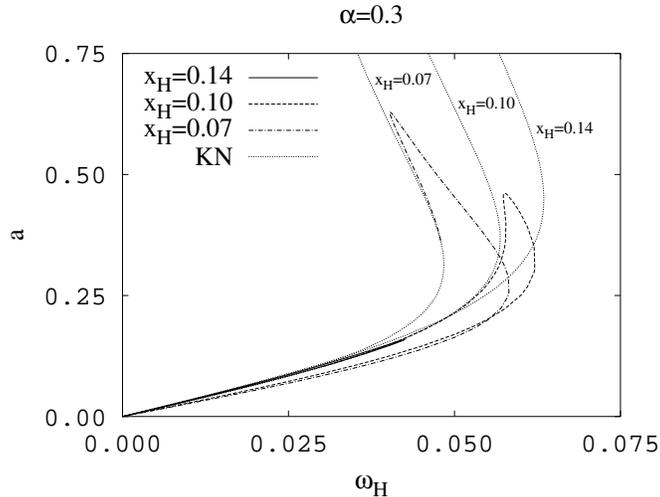}
} } }
\caption{
{\small
Same as Fig.~\ref{f-2} for the specific angular momentum.
}}
\end{figure}

In Fig.~\ref{f-7} we illustrate the surface gravity
of these rotating non-Abelian black holes.
The surface gravity of the non-Abelian black holes shown
is finite. On the other hand,
by keeping the horizon angular velocity $\Omega$ fixed,
and varying $\omega_{\rm H}$, we reach extremal rotating non-Abelian
black holes in the limit $\omega_{\rm H} \rightarrow 0$.
For such extremal black hole the surface gravity vanishes 
\cite{kkrot,kkrotd,long}.

\begin{figure}[h!]
\lbfig{f-7}
\parbox{\textwidth}
{\centerline{
\mbox{
\epsfysize=8.0cm
\includegraphics[width=90mm,angle=0,keepaspectratio]{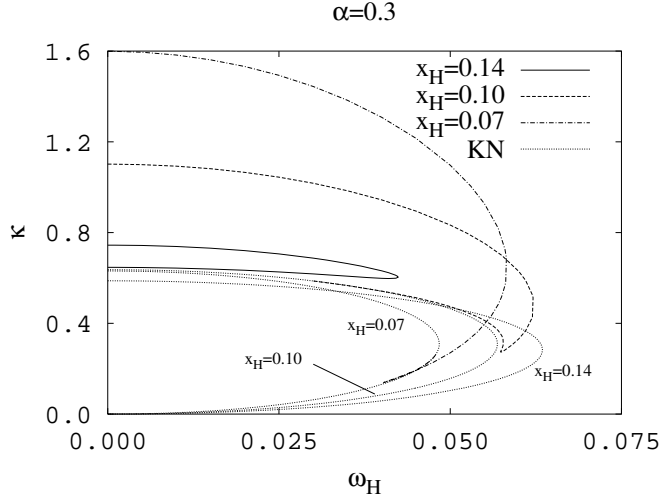}
} } }
\caption{
{\small
Same as Fig.~\ref{f-2} for the surface gravity.
}}
\end{figure}

We finally turn to the gyroelectric ratio $g_{\rm el}$ of the black holes.
Calculating $g_{\rm el}$ for fixed horizon radius $x_{\rm H}$
as a function of the horizon angular velocity parameter $\omega_{\rm H}$
and taking the limit $\omega_{\rm H} \rightarrow 0$,
we extract the gyroelectric ratio $g_{\rm el}$ for 
static non-Abelian black holes, shown in Fig.~\ref{f-8}.
Interestingly, as we approch the static non-Abelian monopole,
by taking the limit $x_{\rm H} \rightarrow 0$,
we find for the gravitating monopole the limiting value $g_{\rm el}=2$,
for all values of $\alpha$ considered.
We note, that the non-Abelian monopole in flat space
also possesses $g_{\rm el}=2$ \cite{gel}, as shown by other means \cite{gel}.

\begin{figure}[h!]
\lbfig{f-8}
\parbox{\textwidth}
{\centerline{
\mbox{
\epsfysize=8.0cm
\includegraphics[width=90mm,angle=0,keepaspectratio]{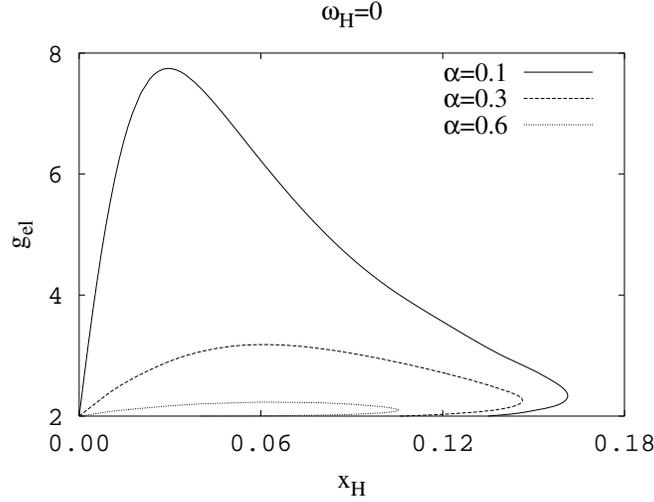}
} } }
\caption{
{\small
The gyroelectric ratio $g_{\rm el}$ versus the
horizon radius $x_{\rm H}$
of non-Abelian black holes, obtained in the limit 
$\omega_{\rm H} \rightarrow 0$
for coupling constant $\alpha=0.1$, 0.3, 0.6.
The limit $x_{\rm H} \rightarrow 0$ yields $g_{\rm el}=2$
for the gravitating monopole.
}}
\end{figure}

Further details will be given elsewhere \cite{long}.

\section{Conclusion}

We have presented the first set of rotating non-Abelian black holes 
of SU(2) EYMH theory.
These black hole solutions are asymptotically flat,
they possess regular horizons, and they carry mass, angular momentum,
and unit magnetic charge. 
In contrast to rotating EYM black holes, no electric
charge is induced by the rotation \cite{vs,kkrot}.
Consequently the gauge fields show a simple power law 
decay at infinity \cite{kkrot,long}.
Distinct from embedded Kerr-Newman solutions,
they represent rotating black hole solutions
with monopole hair.

As in Abelian solutions, 
the rotation induces an electric dipole moment.
Consequently a gyroelectric ratio
can be extracted for these non-Abelian black holes.
This gyroelectric ratio is generically greater than two, the Dirac value
realized in KN black holes.

While non-Abelian monopoles cannot rotate \cite{radu},
we can extract a gyroelectric ratio for gravitating monopoles
by taking the limit 
$\omega_{\rm H} \rightarrow 0$, and $x_{\rm H} \rightarrow 0$
for the rotating non-Abelian black hole solutions.
We thus deduce that gravitating monopoles have a
gyroelectric ratio of two, the value obtained for flat space monopoles
by different means \cite{gel}.

Concerning stability of these rotating non-Abelian black holes,
we expect that, in a certain region of parameter space,
they will be stable, like their static counterparts \cite{gmono1,gmono2,holl}.
Embedded KN black holes should consequently be unstable
in the corresponding region of parameter space \cite{gmono1,gmono2,ewein}.

Extension of this work will include rotating black holes
with higher magnetic charges \cite{kkrotd,hkk},
dyonic rotating black holes \cite{bhk},
and the coupling to a dilaton \cite{dil}.
The latter includes the generalization of the non-Abelian mass formula 
and the uniqueness conjecture to SU(2) EYMH theory \cite{kkrotd}, 
as well as the search for rotating 
black holes with counterrotating horizons \cite{emd}.

\end{document}